\documentclass[preprintnumbers,eqsecnum,prd,showpacs,nofootinbib]{revtex4}
\usepackage{latexsym,amsmath,amssymb}
\usepackage[dvips]{graphicx,color}
\usepackage{dcolumn,bm}    
\newcommand{\ma}[1]{{\mathrm{#1}}}
\newcommand{\calH}{{\cal H}}
\newcommand{\calO}{{\cal O}}
\newcommand{\calN}{{\cal N}}
\newcommand{\calR}{{\cal R}}
\newcommand{\pa}{{\partial}}
\newcommand{\frR}{{\frak R}}

\newcommand{\dd}{{\rm d}}
\newcommand{\vep}{\varepsilon}

\begin{document}
\thispagestyle{empty}


\title{Multi-disformal invariance of nonlinear primordial perturbations}

\preprint{RESCEU-7/15 \,, YITP-15-23}

\author{Yuki Watanabe$^{1}$}
\email[Email : ]{watanabe_at_resceu.s.u-tokyo.ac.jp}
\author{Atsushi Naruko$^{2}$}
\author{Misao Sasaki$^{3}$}

\affiliation{
$^1$ Research Center for the Early Universe, University of Tokyo, Tokyo 113-0033, Japan; \\
Department of Physics, Gunma National College of Technology, Gunma 371-8530, Japan\\
$^2$ Department of Physics, Tokyo Institute of Technology, Tokyo 152-8551, Japan \\
$^3$ Yukawa Institute for Theoretical Physics
 Kyoto University, Kyoto 606-8502, Japan
}

\date{\today}

\begin{abstract}
We study disformal transformations of the metric in the cosmological context.
We first consider the disformal transformation generated by a 
scalar field $\phi$ and show that the curvature and tensor perturbations
on the uniform $\phi$ slicing, on which the scalar field is 
homogeneous, are non-linearly invariant under the disformal transformation.
Then we discuss the transformation properties of the evolution equations 
for the curvature and tensor perturbations at full non-linear order 
in the context of spatial gradient expansion as well as at linear order. 
In particular, we show that the transformation can be described in 
two typically different ways: one that clearly shows the physical invariance
and the other that shows an apparent change of the causal structure.
Finally we consider a new type of disformal transformation in which a
multi-component scalar field comes into play, which we call
a ``multi-disformal transformation''. We show that 
the curvature and tensor perturbations are invariant at linear order,
and also at non-linear order provided that the system has reached
the adiabatic limit.
\end{abstract}

\maketitle

\section{Introduction}
\label{sec:intro}
Cosmic inflation~\cite{Starobinsky:1980te,Sato:1980yn,Guth:1980zm,Linde:1981mu,Albrecht:1982wi} is now a widely accepted paradigm of the very early Universe. 
It provides a mechanism of generating primordial scalar and tensor 
perturbations, both of which are almost scale-invariant, adiabatic, and 
Gaussian: The quantum vacuum fluctuations of a scalar field
and those of gravitons on microscopic scales inside the causal horizon 
({\it i.e.} on scales smaller than the Hubble horizon radius) are 
amplified and stretched to super-horizon scales by the inflationary 
expansion of the Universe, leading to primordial scalar and tensor 
perturbations, respectively. The scalar perturbation is equivalent to
the so-called comoving curvature perturbation (which we call
 simply the curvature perturbation in the rest of this paper)
since it describes the perturbation in the spatial curvature at linear order 
in the comoving slicing.\footnote{The comoving slicing is defined as a slicing 
on which the hypersurface orthogonal vector coincides with the
timelike eigen-vector of the energy momentum tensor (four-velocity in the
case of a fluid), under the assumption that the vector-type perturbation
is negligible.}
The curvature perturbation develops into density inhomogeneities
after it re-enters the Hubble horizon, and seeds the formation of galaxies 
and galactic clusters~\cite{Mukhanov:1981xt}.
The tensor perturbation gives rises to a gravitational 
wave background~\cite{Grishchuk:1974ny, Starobinsky:1979ty,Watanabe:2006qe}. 

As for the curvature perturbation, the above mentioned features
have been confirmed by the full sky observations of 
cosmic microwave background (CMB) anisotropies~\cite{Komatsu:2014ioa, Ade:2013uln}.
In particular, the anti-correlation of the temperature and 
E-mode polarization detected on angular scales of 
$50 \lesssim \ell \lesssim 200 $~\cite{Peiris:2003ff, Komatsu:2014ioa} 
strongly supports the presence of the super-horizon scale curvature perturbation
at the time of decoupling of CMB photons ($z \approx 1090$). 
Moreover, detection of the B-mode polarization on angular scales of 
$ \ell \lesssim 150$, which may become possible in the near future,
 will imply the presence of the super-horizon scale tensor perturbation 
at the decoupling time, and will give further strong evidence for inflation.

Since present and planned-future cosmological observations, including
those of the galaxy distribution, CMB and 21-cm fluctuations on various 
scales are getting precise enough to probe non-linear features such as the
 bispectra and trispectra of scalar and tensor perturbations, 
there is a possibility to identify or constrain the origin of these 
perturbations observationally.
Although all of the single-field slow-roll inflation models with a canonical 
kinetic term predict unobservably small non-linear effects on the primordial 
fluctuations~\cite{Salopek:1990jq,Maldacena:2002vr, Acquaviva:2002ud}
 (see~\cite{Germani:2011ua} for a gravitational enhanced friction model),
any deviation from canonical single-field slow-roll inflation can predict
 detectably large non-Gaussianity of the curvature perturbation, including 
models with non-canonical kinetic terms~\cite{Seery:2005wm, Chen:2006nt, Mizuno:2010ag,Burrage:2010cu, DeFelice:2011zh, Gao:2012ib, Fasiello:2014aqa},
plural light fields~\cite{Suyama:2010uj}, {\it e.g.}, curvaton 
model~\cite{Linde:1996gt, Enqvist:2001zp, Lyth:2001nq, Moroi:2001ct, Sasaki:2006kq},
modulated reheating~\cite{Dvali:2003em, Zaldarriaga:2003my},
and multi-brid inflation~\cite{Sasaki:2008uc}.

Among those non-canonical models, many of them are constructed in 
extended scalar-tensor theories of gravity, such as generalized 
G-inflation~\cite{Kobayashi:2011nu, Deffayet:2011gz, Charmousis:2011bf, Horndeski:1974wa} and its extensions~\cite{Gleyzes:2014dya, Gao:2014soa, Lin:2014jga, Padilla:2012dx, Kobayashi:2013ina}.  
Therefore, studies of non-linear cosmological perturbations within the 
framework of extended gravity are necessary for distinguishing models of inflation.
Such studies are also equally important for testing alternative theories of 
inflation, such as bouncing cosmology~\cite{Xue:2013bva}.  

A formalism has been developed on the basis of spatial gradient expansion
 to study the classical evolution of the curvature perturbation to full
non-linear order on super-horizon scales. 
The zeroth order truncation of gradient expansion corresponds to the 
separate universe approach~\cite{Salopek:1990jq, Sasaki:1998ug, Wands:2000dp}, 
which is the basis of the $\delta N$ 
formalism~\cite{Starobinsky:1986fxa,Sasaki:1995aw, Sasaki:1998ug} that allows 
us to study the nonlinear evolution of the curvature perturbation.
Moreover, at the next leading order, namely the second order
 in the gradient expansion we can take into account
 the effect of a mode closely related to the cosmic shear, which 
usually decays rapidly on super-horizon scales, but which may become important 
in a model where the slow-roll evolution may be violated.
Studies on gradient expansion to second order were carried out for  
single-field k-essence~\cite{Takamizu:2010xy},
for multi-field k-essence~\cite{Naruko:2012fe},
and for the kinetic gravity braiding (KGB) model~\cite{Takamizu:2013gy}.

Creminelli {\it et al.}~\cite{Creminelli:2014wna} have recently shown that
 the primordial linear tensor power spectrum from inflation can be
 always cast into the standard form, i.e., 
$P_\gamma(k) = H^2/(2M_\ma{Pl}^2 k^3)$,
 at leading order in derivatives
 with suitable conformal and disformal transformations
 in the context of the effective field theory (EFT) of
inflation~\cite{Cheung:2007st} where the speed of propagation of 
gravitational waves is not necessarily unity
 and hence the theory of gravity can be modified from general relativity (GR).

Thus, regardless of the theory of gravity, we can formally obtain
 a prediction for linear tensor modes, which strongly indicates
 a close relation between modifying gravity and a disformal plus conformal 
transformation from GR.
Moreover, invariance of the curvature perturbation under disformal transformations
has been shown at linear order~\cite{Minamitsuji:2014waa, Tsujikawa:2014uza}. 

In this paper, we extend the invariance of the curvature and
tensor perturbations to fully non-linear order.
In Sec.~II, we show the invariance under the disformal transformation
to fully non-linear order by exploiting the effect of the transformation
on the form of the metric. We then consider the evolution equations
for the curvature and tensor perturbations, and show explicitly that
the invariance holds not only in linear perturbation theory but also 
full non-linearly up through second order in spatial gradient expansion.
In Sec.~III, we consider a new type of disformal transformation, 
dubbed multi-disformal transformation, generated by 
a multi-component scalar field.
We show that the curvature and tensor perturbations are still invariant
under the multi-disformal transformation at linear order, and
the curvature and tensor perturbations remain invariant again up 
through the second order in gradient expansion,
provided that the system has reached the adiabatic limit.
In doing so, we also show that the curvature and tensor perturbations
in a wide class of extended scalar-tensor theories can be described
in terms of those in GR with redefinitions of background 
quantities.

\section{Invariance of non-linear perturbations and equations}
\label{sec:inv}

\subsection{Invariance of non-linear perturbations}
Let us employ the Arnowitt-Deser-Misner (ADM) formalism and
 the metric is expressed as 
\begin{gather}
 \dd s^2 = g_{\mu\nu} \dd x^{\mu} \dd x^{\nu}
 = - \alpha^2 \dd t^2 + \hat{\gamma}_{i j} \bigl( \dd x^i + \beta^i \dd t \bigr)
 \bigl( \dd x^j + \beta^j \dd t \bigr) \,,
\end{gather}
where $\alpha$ is the lapse function, $\beta^i$ is the shift vector and
 Latin indices run over 1, 2 and 3. 
In addition to the standard ADM decomposition, 
 the spatial metric are further decomposed
 so as to separate trace and unimodular parts as
\begin{align}
 \hat{\gamma}_{i j} &= a^2(t) e^{2\psi} \gamma_{i j} \, , \qquad
 \ma{det} \, \gamma_{i j} = 1 \,, 
\end{align}
where $a(t)$ is the scale factor of a fiducial flat 
Friedmann-Lema\^itre-Robertson-Walker (FLRW) spacetime with the metric,
\begin{align}
 \dd s^2 = - \dd t^2 + a^2 (t) \delta_{i j} \dd x^i \dd x^j \,.
\end{align}
It is sometimes convenient to use the conformal time $\eta$,
 \begin{align}
 \dd \eta = \frac{\dd t}{a (t)} \,,
 \end{align}
in place of the cosmic proper time $t$. Below we denote the proper Hubble
expansion rate of the background FLRW universe by $H$ and the conformal
Hubble expansion rate by $\calH$. That is, $H=\dot a/a$ and $\calH=a'/a$,
where a dot denotes $\dd/\dd t$ and a prime $\dd/\dd\eta$.

The general form of disformal transformation is defined
 as~\cite{Bekenstein:1992pj}
\begin{align}
 \tilde{g}_{\mu \nu} = A (\phi \,, X) \, g_{\mu \nu}
 + B (\phi \,, X) \, \pa_\mu \phi \pa_\nu \phi \, ,\qquad
 X \equiv - g^{\mu \nu} \pa_\mu \phi \pa_\nu \phi/2 \ .
\label{generaldis}
\end{align}
To focus on the effect of disformal transformation, we concentrate
 on the case $A = 1$, that is, a pure disformal transformation.
For simplicity, we simply call it the disformal transformation
throughout this paper unless confusion may arise.
As clear from the above form, the general disformal transformation
consists of a disformal transformation ($A=1$, $B\neq0$) and 
a conformal transformation ($A\neq 1$, $B=0$). 
In the end of this subsection, we also discuss 
the invariance under the conformal transformation.

As a choice of the time slicing, we take the uniform $\phi$ slicing
 on which the scalar field is given as a function of time only, 
that is,
\begin{align}
 \phi = \phi (t) \,.
\end{align}
Under the assumption that the scalar field $\phi$ dominates the universe, 
this slicing is equivalent to the comoving slicing. 
We denote the scalar component of $\gamma_{i j}$ by $\chi$, 
which is extracted from $\gamma_{i j}$ as
 \begin{align}
 \chi \equiv - \frac{3}{4} \, \triangle^{-1} \Bigl\{ \pa^i \Bigl[ e^{- 3 \psi}
 \pa^j \Bigl( e^{3 \psi} (\gamma_{i j} - \delta_{i j}) \Bigr) \Bigr] \Bigr\} \,,
 \end{align}
where $\triangle$ is the flat 3-dimensional Laplacian and 
$\pa^i=\delta^{ij}\pa_j$, and denote $\psi+\chi/3$
on the comoving slicing by $\frR_c$ ($\frR_c \equiv \psi_c + \chi_c/3$)
and its linearized version by $\calR_c$
Namely, $\calR_c$ is the (comoving) curvature
perturbation at linear order and $\frR_c$ its non-linear generalization,
which is the variable of our interest.\footnote{We note that
the non-linear generalization of the curvature perturbation is 
not unique, but it is unique up through second order in spatial
gradient expansion~\cite{Takamizu:2010xy,Takamizu:2013gy}.}
As for the tensor perturbation, 
we define its non-linear generalization as the transverse
part of $\gamma_{ij}$ with respect to the flat background metric 
$\delta_{ij}$. We denote it by $\gamma^{\rm TT}_{ij}$. 
That is, $\partial^j\gamma^{\rm TT}_{ij}=0$.\footnote{Note that 
$\gamma^{\rm TT}_{ij}$ is traceless only at linear order, in spite of 
the fact that we use the superscripts ${\rm TT}$ for convenience, 
which could mean transverse and traceless.}
We note that $\gamma^{\rm TT}_{ij}$ is independent of the
time-slicing condition at linear order but is slice-dependent
at higher orders. Thus at non-linear order our discussion applies
only to $\gamma^{\rm TT}_{ij}$ defined on the uniform $\phi$ slicing.

An immediate consequence here is that the disformal transformation 
 affects only the $(0 \,, 0)$-component of the metric:
\begin{align}\label{eq:disformal}
 \tilde{g}_{\mu \nu} = g_{\mu \nu}
 + B \, \dot{\phi}^2 \, \delta_\mu{}^0 \delta_\nu{}^0 \,,
\end{align}
 and hence only the lapse function,
\begin{align}
 \tilde{\alpha}^2 = \alpha^2 - B \dot{\phi}^2 \,.
\label{alphatrans}
\end{align}
For completeness, let us write down the transformed metric,
\begin{align}
\dd \tilde{s}^2 = \tilde{g}_{\mu\nu} \dd x^{\mu} \dd x^{\nu}
 = - \tilde{\alpha}^2 \dd t^2 
+ \hat{\gamma}_{i j} \bigl( \dd x^i + \beta^i \dd t \bigr)
 \bigl( \dd x^j + \beta^j \dd t \bigr) \,,
\end{align}
and the corresponding FLRW background,
\begin{align}
\dd \tilde{s}^2 = -\tilde{\alpha}_0^2 \dd{t}^2
 + a^2 (t) \delta_{i j} \dd x^i \dd x^j \,,
\end{align}
where $\tilde{\alpha}_0$ is the background value of $\tilde{\alpha}$.

Thus, with the choice of the uniform $\phi$ slicing,
the disformal transformation only induces the change in
the lapse function and hence the spatial metric as well as the shift vector are
 trivially invariant to fully non-linear order.
This fact implies the invariance of the curvature perturbation 
 on the uniform $\phi$ slicing as well as that of the tensor perturbation
 which is contained in the unimodular part of the spatial metric.

Next let us discuss the invariance of the curvature and tensor
perturbations under the conformal transformation.
It is widely known that when $A$ in (\ref{generaldis})
is a function of $\phi$ only,
 the curvature and tensor perturbations on the uniform $\phi$ slicing
 are invariant even at non-linear level~\cite{Gong:2011qe}.
Here we shall extend their discussion by introducing the dependence 
on $X$ in $A$.
First as for the tensor perturbation, it is manifestly invariant
 even under this extended conformal transformation
 because the conformal factor does not affect the unimodular part
 of the spatial metric.
On the other hand, the curvature perturbation is no more invariant under 
this transformation in general.
By perturbatively expanding the transformation equation,
\begin{align}
 \tilde{g}_{\mu \nu} = A \, g_{\mu \nu}
 = \bar{A} \, g_{\mu \nu} + \delta A \, \bar{g}_{\mu \nu} 
 + \delta A \, \delta g_{\mu \nu} \, ,
\end{align}
 one easily notices that the second (as well as third) term on the RHS
 implies the change in the curvature perturbation as
$\calR \to \calR + \delta A/{2} + \cdots$, 
 while the first term only affect the (background) scale factor. 
However there is a case where the invariance of curvature perturbation still
 holds, namely when $\delta A$ vanishes.
Since in the uniform $\phi$ slicing $\phi$ is a function of time,
 $\delta A$ is sourced by the perturbation of $X$, which in turn comes from
 that of the lapse function because
 \begin{align}
 X_c = \frac{1}{2} \frac{ \dot{\phi}^2 (t)}{\alpha_c^2 (t \,, {\bm x})}
 = \frac{1}{2} \dot{\phi}^2 (t)
 - \delta \alpha_c (t \,, {\bm x}) \dot{\phi}^2 (t) + \cdots \,,
 \end{align}
where the subscript $c$ is for the comoving (or uniform $\phi$) slicing.
In the context of single-field inflation, it is well known that
 the curvature perturbation on the uniform $\phi$ slicing is
 conserved on large scales.\footnote{Strictly speaking, 
the argument cannot be applied to the so-called ultra 
slow-roll model~\cite{Namjoo:2012aa,Martin:2012pe}.}
This fact may be regarded as a consequence of
the vanishing of the lapse function perturbation
 on large scales, $\delta\alpha_c=\calO (\vep^2)$ where
$\vep$ represents terms of first order in spatial derivatives,
because the evolution of the curvature perturbation is
 solely induced by the lapse function in the uniform $\phi$ slicing,
 $\dot{\calR}_c \propto H \delta \alpha_c$,
 which is valid not only in GR but also in
Horndeski~\cite{Kobayashi:2011nu, Deffayet:2011gz, Charmousis:2011bf, Horndeski:1974wa}
 and GLPV~\cite{Gleyzes:2014dya} theories.

To summarize we conclude that the invariance of the curvature perturbation
holds under the pure disformal transformation ($A=1$) to full non-linear
order, and is effectively realized to full non-linear order on superhorizon 
scales (or at leading order in gradient expansion) under the most
general disformal transformation given by (\ref{generaldis}),
while the tensor perturbation defined on the comoving slicing is
invariant under the most general disformal transformation to full
 non-linear order.

\subsection{Transformation of non-linear equations}
Here the effects of the disformal transformation on
the evolution equations for the scalar and tensor perturbations
are discussed for linear theory and for fully non-linear theory
but in the context of gradient expansion.

In linear theory, the equation for $\calR_c$ in the case of
 a canonical scalar field in GR is
\begin{align}\label{eq:calR-linear}
 \frac{1}{z^2}\frac{1}{\alpha_0}\frac{\dd}{\dd\eta}
\left(\frac{z^2}{\alpha_0}\frac{\dd}{\dd\eta}\calR_c \right)
+ c_s^2k^2 \calR_c = 0 \,, \quad
 z \equiv a \frac{\phi'}{\calH} \,,
\end{align}
where ${\alpha}_0$ is the background value of the lapse function and
 $c_s$ is the sound velocity, both of which are unity
in the present case, $\alpha_0=1$ and $c_s=1$. 
However, we keep them here for convenience.

Noting (\ref{alphatrans}), the disformal transformation gives 
\begin{align}
 \frac{1}{z^2}\frac{1}{\tilde{\alpha}_0}\frac{\dd}{\dd\eta}
\left(\frac{z^2}{\tilde{\alpha}_0}\frac{\dd}{\dd\eta}\calR_c \right)
+ c_s^2k^2 \calR_c = 0 \,, \quad
 z \equiv a \frac{\phi'}{\calH} \,.
\label{eq:calR-linear-dis}
\end{align}
Comparing (\ref{eq:calR-linear}) and (\ref{eq:calR-linear-dis}),
it is readily apparent that both of them are in the same form if
the derivatives are expressed in terms of the proper time defined 
in respective frames, 
$\dd\tau=\alpha_0\dd t$ and $\dd\tilde{\tau}=\tilde{\alpha}_0\dd t$
($\dd t=a\dd\eta$).
There is no change in $z$ nor in $c_s$.

On the other hand, sometimes it is convenient to
 write the transformed equation by redefining $z$ and $c_s$,
\begin{align}
 \frac{1}{\tilde{z}^2}\frac{1}{\alpha_0}\frac{\dd}{\dd\eta}
\left(\frac{\tilde{z}^2}{\alpha_0}\frac{\dd}{\dd\eta}\calR_c \right)
+ \tilde{c}_s^2k^2 \calR_c = 0 \,, \quad
 z \equiv a \frac{\phi'}{\calH} \,,
\end{align}
where we have defined
\begin{align} \tilde{c}_s \equiv \tilde{\alpha}_0c_s \,, \qquad
 \tilde{z}  \equiv \frac{z}{\sqrt{\tilde{\alpha}_0}} \,.
\label{eq:redef_cs}
\end{align} 
With this redefinition of the background quantities, the equation for 
$\calR_c$ can be reinterpreted as the one in a modified theory of gravity.
Conversely, this implies that the equation for the curvature perturbation 
in a theory of modified gravity may be put in the form of the one in GR by
a suitable disformal transformation.

Now we turn to the non-linear case.
In the context of spatial gradient expansion,
 the full non-linear equation for $\frR_c$ is given by~\cite{Takamizu:2010xy}
\begin{align}
  \frac{1}{z^2}\frac{1}{{\alpha}_0}\frac{\pa}{\pa \eta}
\left(\frac{z^2}{{\alpha}_0}\frac{\pa}{\pa \eta}\frR_c\right)
  + \frac{c_s^2}{4}\, 
{}^{(3)}R\bigl[e^{2\psi}\gamma_{ij}\bigr] = \calO (\vep^4) \,,
\end{align}
with the same $z$ as in \eqref{eq:calR-linear}.
Here $\psi=\frR_c+\calO(\epsilon^2)$ and 
${}^{(3)}R$ is the spatial scalar curvature.
 It was later extended so as to include the KGB term in \cite{Takamizu:2013gy}.
 By the same reasoning as in the linear case, the nonlinear equation for $\frR_c$ 
remains invariant if expressed in terms of the proper time in each frame,
but it can also be interpreted as the one in a modified theory of gravity
with modified $z$ and $c_s$ as defined by \eqref{eq:redef_cs}.
 
 As for the tensor perturbation, one can find the full-nonlinear equation
 for $\gamma^{TT}_{i j}$: 
 \begin{align}
\frac{1}{z_t^2}\frac{1}{\alpha_0}
\frac{\pa}{\pa \eta}\left(\frac{z_t^2}{\alpha_0}\frac{\pa}{\pa \eta}
\gamma^{\rm TT}_{i j}\right)
 + \frac{1}{4} \Bigl(e^{- 2 \psi}\, {}^{(3)}R_{i j} \bigl[ e^{2 \psi}
 \gamma_{i j} \bigr] \Bigr)^{\rm TT} = \calO (\vep^4) \,,
 \end{align}
 where $z_t\equiv a$ and $(\cdots)^{\rm TT}$ denotes 
the transverse-traceless projection,\footnote{Note that
the time derivative of $\gamma^{\rm TT}_{ij}$ is of second order
in gradient expansion and is traceless, though 
$\gamma^{\rm TT}_{ij}$ itself is not so.}
$\psi=\frR_c+\calO(\epsilon^2)$ and ${}^{(3)}R_{ij}$ is the spatial Ricci tensor.
Thus the same argument as the scalar case applies to the 
tensor case as well. Namely, the equation takes exactly the same form
if expressed in terms of the proper time. Also, the equation
after the disformal transformation may be expressed as the one
in a modified gravity theory,
 \begin{align}
\frac{1}{\tilde{z}_t^2}\frac{1}{\alpha_0}
\frac{\pa}{\pa \eta}\left(\frac{\tilde{z}_t^2}{\alpha_0}\frac{\pa}{\pa \eta}
\gamma^{\rm TT}_{i j}\right)
 + \frac{\tilde{c}_t^2}{4} \Bigl(e^{- 2 \psi}\,{}^{(3)}R_{i j} \bigl[ e^{2 \psi}
 \gamma_{i j} \bigr] \Bigr)^{\rm TT} = \calO (\vep^4) \,,
 \end{align}
where we have defined
\begin{align}
\tilde{c}_t \equiv \tilde{\alpha} \,, \qquad
\tilde{z}_t \equiv \frac{z_t}{\sqrt{\tilde{\alpha}}}
= \frac{a}{\sqrt{\tilde{\alpha}}}  \,.
\end{align}

To conclude this subsection, we have explicitly shown that 
the evolution equations for the curvature and tensor perturbations
are invariant under the disformal transformation
not only in linear perturbation theory but also non-linearly 
up through second order in gradient expansion. We have
also shown that the transformed equations can also be cast in the 
form of those in a theory of modified gravity.

\section{Multi-disformal transformation}
\label{sec:multi}

In this section, we consider a new type of disformal
transformation generated by a multi-component scalar field,
dubbed ``multi-disformal transformation''.
Suppose there are $\calN$ component scalar field,
$\phi^I \, (I = 1 , \cdots \,, \calN)$.
Extending the general disformal transformation, we consider
\begin{align}
 \tilde{g}_{\mu \nu} = A (\phi^I \,, X^{IJ}) \, g_{\mu \nu}
 + B_{KL} (\phi^I \,, X^{IJ}) \, \pa_\mu \phi^K \pa_\nu \phi^L
\end{align}
 where
 \begin{align}
 X^{IJ} = - \frac{1}{2}g^{\mu \nu} \pa_\mu \phi^I \pa_\nu \phi^J\,. 
 \end{align}
Again we focus on the pure disformal part, $A = 1$.
An immediate consequence is that the linear curvature and tensor perturbations
 are invariant even in this general transformation because
 the contribution to the spatial section of the metric is of 
second order in perturbation, i.e., 
$B_{IJ} \pa_i \phi^I \pa_j \phi^J = \calO (\delta^2)$ where
$\delta$ represents the order of linear perturbation. 
Furthermore because it is also of second order in spatial derivatives,
the curvature and tensor perturbations are invariant to full non-linear
order at leading order in gradient expansion.
However apparently this result cannot be extended to second order
in gradient expansion, nor to second order in perturbation.

Nevertheless, there exists a situation where the non-linear
invariance still holds, namely, when the system has converged to 
a state where all the components are determined by a unique scalar 
function, say $\varphi$. Conventionally it is said that a system has 
reached the adiabatic limit when such a state is realized.
Now let us consider the adiabatic-limit,
 \begin{align}
 \phi^I = \phi^I (\varphi)  .
 \end{align}
Restricting to $A =1$, the above transformation in this limit 
reduces to 
 \begin{align}
 \tilde{g}_{\mu \nu} = g_{\mu \nu}
 + B_{KL} \Bigl[ \phi^I (\varphi) \,, 
X^{IJ} (\varphi\,, \pa \varphi) \Bigr] \, (\phi^K)' (\phi^L)' 
\pa_\mu {\varphi} \pa_\nu{\varphi} \,, \qquad
 (\phi^I)' \equiv \frac{\pa \phi^I}{\pa\varphi}\,.
 \end{align}
Here a prime denotes $\dd/\dd\varphi$, not to be confused with 
the conformal time derivative.
Then by taking a uniform $\varphi$ slicing, this equation further 
reduces to
 \begin{align}
 \tilde{g}_{\mu \nu} = g_{\mu \nu}
 + B_{KL} \Bigl[ \phi^I (\varphi) \,, 
X^{IJ} (\varphi \,, \dot{\varphi}/\alpha) \Bigr]
 \, (\phi^K)' (\phi^L)' \dot{\varphi}^2\delta_\mu{}^0 \delta_\nu{}^0 \,.
 \end{align}
Since the disformal transformation only affects the lapse function,
we can apply the same argument as in Sec.~\ref{sec:inv}
and conclude again that the curvature and tensor perturbations
defined on the uniform $\varphi$ slicing
are invariant under the disformal transformation to full non-linear order.

Finally, as for the conformal transformation, the invariance holds
in the adiabatic limit if $A$ is a function of only $\phi^I$.
The invariance does not hold in general if $A$ contains $X^{IJ}$ dependence. 
Nevertheless, if we focus on the adiabatic limit, 
we can again apply the same argument as in Sec.~\ref{sec:inv} and 
show the invariance on superhorizon scales at leading order in
gradient expansion.

\section{Summary}
\label{sec:sum}

In this paper we have discussed the disformal transformation 
in the context of cosmology, particularly of inflationary cosmology.
We have shown that the curvature and tensor perturbations on the 
uniform $\phi$ slicing are fully non-linearly
invariant under the disformal transformation by exploiting the
form of the transformed metric.
We have also shown explicitly to second order in gradient expansion
that the evolution equations for the curvature and tensor fluctuations 
are invariant under the disformal transformation.
In doing so, we have shown another aspect of the disformal
transformation, namely the transformed equations may be
expressed in the form which are identical to those in a modified 
gravity theory.

Then we have discussed a new type of disformal transformation,
 dubbed ``multi-disformal transformation,'' which is an extension of
 the disformal transformation to the multi-component field case.
In this case assuming the system has reached the adiabatic limit
where all the components of the scalar field depend only on
 a single scalar function, $\varphi$, we have also shown the full 
nonlinear invariance of the curvature and tensor perturbations 
on the uniform $\varphi$ slicing.

It seems there are several important or interesting applications 
of the results obtained in this paper.
First as we have seen, utilizing disformal and conformal transformations
 one can modify the background quantities appearing in the perturbation
equations such as the speed of propagation of a perturbation.
Thanks to this property, once a second order differential equation is
 obtained either from a concrete theory like Galileon theory or EFT approach
to inflation, one can map it to the form same as the one in GR by
a suitable disformal transformation without solving the equations.
We have shown that this property can be extended to non-linear equations 
up to second order in gradient expansion. In fact, the full nonlinear 
invariance of the curvature and tensor perturbations suggests that
this mapping may be extended to full non-linear order without resorting
to gradient expansion~\cite{Guillem}. This will make a very useful tool 
for studying the cosmological perturbations in otherwise complicated models. 
Definitely further studies in this direction seems fruitful.

As for the multi-disformal transformation, it might play an important role
in uncovering a new theory.
After the (re-)discovery of the generalized Galileon or
Horndeski theory~\cite{Horndeski:1974wa}, which is the 
most general scalar-tensor theory with the second-order field equations, 
it has been further extended to a more general class of 
theories~\cite{Gleyzes:2014dya}. The relations among those theories 
under disformal plus conformal transformations have been partially studied 
in the literature, but they have not been fully understood yet.
The relation between Galileon/Horndeski theory and DBI theory under the disformal 
transformation has been also discussed~\cite{deRham:2010eu,Goon:2011uw}, 
but restricted to a very special case. 

For multi-field extensions of Galileon/Horndeski theory, there are so far
 two attempts in the literature: one is a (not necessarily general)
multi-field extension of the single-field Galileon/Horndeski 
theory~\cite{Padilla:2012dx}, and the other is a multi-scalar 
version of DBI Galileon 
theory~\cite{Hinterbichler:2010xn,RenauxPetel:2011dv,RenauxPetel:2011uk}
obtained by a special form of the multi-disformal transformation
applied to the DBI theory.
It has been found that thus obtained multi-DBI Galileon theory contains
terms which are absent in the above-mentioned multi-field
Galileon/Horndeski theory~\cite{Kobayashi:2013ina}. 
This fact indicates the existence of a more general class of 
multi-scalar-tensor theories with second-order field equations. 
Thus studies of the multi-disformal transformation in its most general form
would shed light on new theoretical possibilities and lead us
to new cosmological models.

\acknowledgments
We would like to thank Hayato Motohashi and Jonathan White for discussions 
on the invariance of the curvature perturbation under disformal 
transformations. We understand that they also have just completed an
article on a similar topic~\cite{HMotoJwhite}. 
We have also benefited substantially from fruitful discussions with 
Guillem Domenech on various aspects of the disformal transformation.
We would also like to thank Jinn-Ouk Gong and Masahide Yamaguchi
 for useful discussions especially during the workshop
 ``Miniworkshop on cosmology''.
A.N. is grateful to the Yukawa Institute for Theoretical Physics
 at Kyoto University for warm hospitality
 where this work was initiated, advanced and completed. 
This work was supported in part by the JSPS Research
 Fellowship for Young Scientists Nos. 269337 (Y.W.) and 263409 (A.N.).


\end{document}